\documentclass[aps,prb,showpacs,groupedaddress,superscriptaddress,twocolumn]{revtex4}
\usepackage{epstopdf}
\usepackage{mathrsfs}
\usepackage{graphicx,amsmath}
\usepackage{color}
\begin{document}

\title{Gate voltage controlled thermoelectric figure of merit in three-dimensional topological insulator nanowires}

\author{Ning-Xuan Yang}
\affiliation{International Center for Quantum Materials, School of Physics, Peking University, Beijing 100871, China}
\affiliation{Collaborative Innovation Center of Quantum Matter, Beijing 100871, China}

\author{Yan-Feng Zhou}
\affiliation{International Center for Quantum Materials, School of Physics, Peking University, Beijing 100871, China}
\affiliation{Collaborative Innovation Center of Quantum Matter, Beijing 100871, China}

\author{Peng Lv}
\affiliation{International Center for Quantum Materials, School of Physics, Peking University, Beijing 100871, China}
\affiliation{Collaborative Innovation Center of Quantum Matter, Beijing 100871, China}

\author{Qing-Feng Sun}
\email[]{sunqf@pku.edu.cn}
\affiliation{International Center for Quantum Materials, School of Physics, Peking University, Beijing 100871, China}
\affiliation{Collaborative Innovation Center of Quantum Matter, Beijing 100871, China}
\affiliation{CAS Center for Excellence in Topological Quantum Computation, University of Chinese Academy of Sciences, Beijing 100190, China}

\date{\today}

\begin{abstract}
The thermoelectric properties of the surface states in three-dimensional
topological insulator nanowires are studied.
The Seebeck coefficients $S_c$ and the dimensionless thermoelectrical figure of merit $ZT$ are obtained
by using the tight-binding Hamiltonian combining with the nonequilibrium Green's function method.
They are strongly
dependent on the gate voltage and the longitudinal and perpendicular magnetic fields.
By changing the gate voltage or magnetic fields, the values of $S_c$ and $ZT$ can be easily controlled.
At the zero magnetic fields and zero gate voltage, or at the large
perpendicular magnetic field and nonzero gate voltage, $ZT$ has the large value.
Owing to the electron-hole symmetry, $S_c$ is an odd function of the Fermi energy
while $ZT$ is an even function regardless of the magnetic fields.
$S_c$ and $ZT$ show peaks when
the quantized transmission coefficient jumps from one plateau to another.
The highest peak appears while the Fermi energy is near the Dirac point.
At the zero perpendicular magnetic field and zero gate voltage,
the height of $n$th peak of $S_C$ is
$\frac{k_B}{e}\texttt{ln}2/(|n|+1/2)$ and $\frac{k_B}{e}\texttt{ln}2/|n|$
for the longitudinal magnetic flux $\phi_{\parallel} = 0 $ and $\pi$, respectively.
Finally, we also study the effect of disorder and find that $S_c$ and $ZT$ are
robust against disorder.
In particular, the large value of $ZT$ can survive even if at the strong disorder.
These characteristics (that $ZT$ has the large value, is easily regulated,
and is robust against the disorder)
are very beneficial for the application of the thermoelectricity.
\end{abstract}


\maketitle

\section{\label{sec1}Introduction}

In recent years, the discovery of the three-dimensional (3D) topological insulators (TIs) has opened up a new
field for the condensed matter physics, which is also one of the most important advances
in material science.\cite{KCL,BBA,KM} TIs have attracted wide attention
because of the exotic physical properties and potential
huge applications in spintronics.\cite{HMZ,Qxl,Zhangliu,Chenyl,Xiay}
TIs are characterized by the insulating bulk states and
nontrivial conducting surface state, which is topologically protected by
time-reversal symmetry. The time reversal invariant disorders can not
cause the backscattering and can not open the gap on the surface states.
The surface states, which present an odd number of gapless
Dirac cones, are featured by the unique Dirac-like linear dispersion
with the spin-momentum locked helical properties.\cite{HD,Xuy1,RYoshimi,NKoirala}
Moreover, for a TI nanowire, a gap opened in the surface states results from the $\pi$ Berry phase
obtained by the 2$\pi$ rotaition of the spin around the nanowire.\cite{JBardarson,REgger,Zhangy}
However, by threading a magnetic flux $\phi_0/2=h/2e$ paralleling the wire, an extra дл Aharonov-Bohm
phase cancels the $\pi$ Berry phase and closes the gap, i.e. wormhole effect.\cite{HPeng,FXiu,Dufouleur,SHong,SCho}

The materials used to make thermoelectric generators or thermoelectric refrigerators
are called thermoelectric materials, which can directly convert the thermal energy into
the electrical energy each other.
Thermoelectric materials have wide application prospects in thermoelectric power generation
and thermoelectric refrigeration. Using thermoelectric materials to generate electricity
and refrigeration will effectively solve the problem of energy sustainable utilization.
TIs share similar material properties, such as heavy elements,
narrow band gaps and quantum localization effect, with thermoelectric materials.
Many TIs (like $\textrm{Bi}_2\textrm{Te}_3$, $\textrm{Sb}_2\textrm{Te}_3$ and $\textrm{Bi}_x\textrm{Sb}_{1-x}$) are
considered as excellent
materials for thermoelectric conversion.\cite{Muchler,XuN,HeJ}
The new physical properties of TIs nanomaterials bring new breakthroughs to the research
of thermoelectric materials and provide new opportunities for the development of thermoelectric
technology. Therefore, it is very important and necessary to find high-efficiency
thermoelectric materials.
The conversion efficiency of thermoelectric materials depends on the dimensionless thermoelectrical figure of merit $ZT$.
$ZT$ is defined as $ZT=\sigma{{S_c}^2}\mathcal{T}/\kappa$, where $\sigma$ is the electric conductivity,
$S_c$ is the Seebeck coefficient, and $\mathcal{T}$ is the operating temperature of the device,
and the thermal conductivity $\kappa$ is the sum of the electric thermal conductivity
and lattice-thermal conductivity.\cite{Muchler,XuN,HeJ,Liuj,Xingy,Weim}
The higher $ZT$ value of thermoelectric material, the better its performance.
There are two ways to raise the $ZT$ value. One is to increase the
thermopower $S_c$ and electrical conductivity. The large thermopower $S_c$
can convert the temperature difference to the voltage at both ends of the material more effectively.
The other is to reduce the thermal conductivity to minimize the energy loss
induced by heat diffusion and Joule heating.
However, due to the restriction of the Mott relation\cite{Cutler} and
the Wiedemann-Franz law,\cite{Jeffrey} a high thermopower $S_c$ leads
to a low electrical conductance, and a high electrical conductivity in a material also implies a high thermal conductivity.
These three parameters need to be optimized to maximize the $ZT$ value.
The study of thermoelectric transport characteristics would be helpful in improving the
conversion efficiency between the electrical energies and the thermal energies.\cite{Muchler,XuN,HeJ,Liuj,Weim}

Generally, we consider the thermoelectric power, also called Seebeck coefficient which
measures the  magnitude of the longitudinal current induced by a longitudinal thermal
gradient in the Seebeck effect.
The thermoelectric power derived from the balance between the electric and thermal forces acting
on the charge carrier, is more sensitive to the details of the density of states
than the electronic conductance.\cite{Xingy,Cutler,Abrikosov,Beenakker,acheng,LiYX}
Therefore, the thermoelectric power is more helpful to
understand the particle-hole asymmetry of TIs.
The thermoelectric power can clarify the details of the electronic structure
of the ambipolar nature for the TI nanowires more clearly than the detection of conductance alone.
Although the Seebeck effect and the Peltier effect provide a theoretical principle
for the application of thermoelectric energy conversion and thermoelectric refrigeration,\cite{Weim,Callen}
the classical Mott relation and the Wiedemann-Franz law may not be established due to the
quantum behavior in nanostructured materials. Therefore, the study of thermoelectric power may
inspires new ideas in the design of quantum thermoelectric devices.\cite{Kubala}

In 1993, Hicks and Dresselhaus\cite{Hicks} found that $ZT$ value increases
swiftly as the dimensions decrease and strongly depends on the wire width.
Hicks and Dresselhaus proposed the idea of using low-dimensional
structural materials to obtain high $ZT$. Then more and more research groups begin
to pay attention to the thermoelectric transport properties
in nanostructure materials.\cite{Miyasato,Onoda,MaR,XuyGan,ZhangFeng,Rameshti,Lijw,Matsushita,Shapiro,Limms}
Especially in recent years, with the development
of the low-temperature measurement technology and the improvement of the
microfabrication technology,
the thermoelectric measurement in low-dimensional samples has became feasible
at low temperature, and various groups were able to fabricate nanostructures
and measure their thermoelectric properties at low
temperature.\cite{Miyasato,ZhangFeng,Matsushita,Shapiro}
In addition, the charge carrier density in nanostructured materials can easily be
tuned globally or locally by varying the magnetic field or the applied gate
voltage. Due to the thermoelectric effect being sensitive to the changes of carrier density, the
$S_c$ and $ZT$ of thermoelectric materials can be controlled by applying
in different directions of the magnetic field
and changing the gate voltage, which opens up a broad way to find
high-efficiency thermoelectric materials.\cite{Liuj}

In this paper, we carry out a theoretical study of the thermoelectric properties of 3D TI nanowires under the longitudinal
and perpendicular magnetic fields by using the Landauer-B\"{u}ttiker formula
combining with the nonequilibrium Green's-function method.
While the Fermi energy just crosses discrete transverse channels,
the transmission coefficient of the quantized plateaus jumps from one step to another
and the Seebeck coefficient $S_c$ and the thermoelectric figure of merit $ZT$ show peaks.
Due to the electron-hole symmetry, $S_c$ is odd function of the Fermi energy $E_F$, and $ZT$ is even function.
$S_c$ and $ZT$ have very large peaks near the Dirac point at the zero magnetic field
and zero gate voltage, because of the extra $\pi$ Berry phase around the TI
nanowire and a gap appearance in the energy spectrum.
The thermoelectric properties of the TI nanowire are obviously dependent on
the gate voltage and the longitudinal and perpendicular magnetic fields.
The values of $S_c$ and $ZT$ can be easily controlled
by changing the gate voltage or magnetic fields.
In addition, the effect of the disorder on the thermoelectric properties is also studied.
The Seebeck coefficient $S_c$ and $ZT$ are robust against the disorder, but the
plateaus in the conductance are broken. This is very counterintuitive. In usual,
the $S_c$ and $ZT$ are more sensitive than the conductance.
In particular, the large peak value of $ZT$ can well survive,
which is very promising for the application of the thermoelectricity.

The rest of the paper is organized as follows.
In Sec.~\ref{sec2}, the effective tight-binding Hamiltonian is introduced.
The formalisms for calculating the Seebeck coefficient $S_c$ and
the thermoelectric figure of merit $ZT$ are then derived.
In Sec.~\ref{sec3}, the thermoelectric properties at zero perpendicular magnetic field
and zero gate voltage are studied.
Sec.~\ref{sec4} and Sec.~\ref{sec5} contribute to the effect of
the perpendicular magnetic field, gate voltage, and disorder on
thermoelectric properties, respectively.
Finally, a brief summary is drawn in Sec.~\ref{sec6}.

\section{\label{sec2}Model and Methods}

Here we consider a cuboid 3D TI nanowire under the longitudinal and perpendicular magnetic fields
as shown in Fig.1(a).
Based on the lattice model, the two-dimensional Hamiltonian for
surface states of the 3D TI nanowire can be described as follows,\cite{Zhouyf}
\begin{eqnarray}\label{eq:1}
H&=&\sum_{m} \left[ \sum^{N}_{n}c_{nm}^{\dag}R_{0}c_{nm}+ \sum^{N}_{n} c_{nm}^{\dag}R_{y}c_{n,m+1} \right.\nonumber\\
&+ & \left. \sum^{N-1}_{n}c_{nm}^{\dag}R_{xn}c_{n+1,m} -c_{Nm}^{\dag}R_{xN}c_{1m}+\textrm{H.c.}\right],
\end{eqnarray}
with
\begin{eqnarray}\label{eq:a2}
R_0& =&(2W/a)\sigma_z+ U_{n}\sigma_0, \nonumber\\
R_{xn}& =&[-(W/2a)\sigma_z+(i\hbar \nu_{F}/2a)\sigma_{y}]e^{i(\phi^{\parallel}_{n,n+1}+\phi^{\perp}_{n,n+1})},\nonumber\\
R_y& =&-(W/2a)\sigma_z-(i\hbar \nu_{F}/2a)\sigma_{x},
\end{eqnarray}
where $c_{nm}$ and $c_{nm}^{\dag}$ are the annihilation and creation operators at site $(n,m)$ respectively,
with the index $m$ being along the y-direction and $n$ being along the circumference of the TI
nanowire.
$N$ is the total number of lattices encircling the TI nanowire,
$a$ is the lattice constant, $\nu_F$ is the Fermi velocity, $\sigma_x,\sigma_y,\sigma_z$ are the Pauli matrices, $\sigma_0$ is the unit matrix,
and $U_{n}$ is the on-site energy which can be regulated by the gate voltage.
Here we set $U_{n} =\Delta U /2$ for the upper surface, $U_{n} =-\Delta U /2$ for the lower surface,
and $U_{n}$ is linear from $\Delta U /2$ to $-\Delta U /2$ for two side surfaces.
Here the effect of longitudinal magnetic field is included by adding a phase term $\phi^{\parallel}_{n,n+1} =\int_{n}^{n+1}{\bf A_{\parallel}}\cdot d{\bf l}/\phi_0$ to $R_{xn}$ in Eq.(\ref{eq:a2}),
where ${\bf A_{\parallel}}=(0,0,B_{\parallel}x)$ is the vector potential for a magnetic field $B_{\parallel}$ parallel to the $y$ direction. Furthermore, we also consider a uniform magnetic field $B_{\perp}$ perpendicular
to the upper and lower surfaces [see Fig.1(a)],
then a phase $\phi^{\perp}_{n,n+1} =\int_{n}^{n+1}{\bf A_{\perp}}\cdot d{\bf l}/\phi_0$ is added
in the hopping term $R_{xn}$, where ${\bf A_{\perp}}=(B_{\perp} y,0,0)$.
$W$ in Eq.(\ref{eq:a2}) is the Wilson term.
The Wilson term is introduced for solving the fermion doubling problem in the lattice model.\cite{Zhouyf}
In the numerical calculations, $W$ is set $0.3\hbar\nu_F$.
The nanowire is assumed to have a cross section of a size $(L_x, L_z)=(96 \ \mathrm{nm}, 12\ \mathrm{nm})$.
For the nanowires of other sizes, the results are similar.
We also set the Fermi velocity $\nu_{F} = 5\times10^{5}\ \rm m/s$
and the lattice constant $a =0.6\ \mathrm{nm}$.\cite{RYoshimi,ZhangT}

\begin{figure}
\includegraphics[scale=0.38]{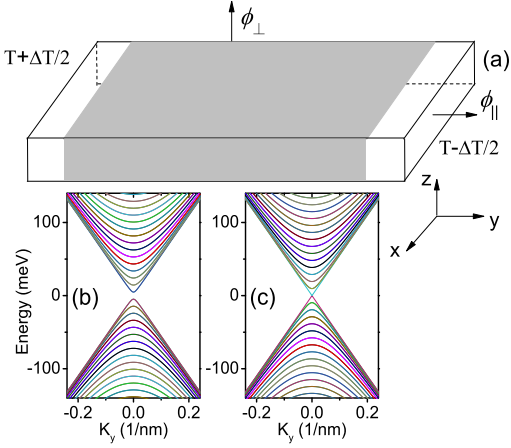}
\caption{ (a) The schematic of a cuboid 3D TI nanowire
under the longitudinal and perpendicular magnetic fields. The grey region is
the center scattering region.
(b) and (c) show the energy band structure of the TI nanowire
with the longitudinal magnetic flux $\phi_{\parallel} = 0$ and $\pi$, respectively.
The perpendicular magnetic field $\phi_{\perp}=0$, gate voltage $\Delta U =0$, and the disorder strength $D=0$.
}
\end{figure}

Considering that the bias and temperature of the left/right terminal are
$V_{L/R}$ and $\mathcal{T}_{L/R}$,
the electronic current and the electric-thermal current flowing from the left terminal to
the cuboid 3D TI nanowire can be calculated from the Landauer-B\"{u}ttiker formula,\cite{Weim}
\begin{equation}\label{eq:2}
\begin{split}
&J_L =\frac{e}{h} \int
T_{\mathrm{LR}}(E)[f_{\mathrm{L}}(E)-f_{\mathrm{R}}(E)]dE,\\
&Q_L =\frac{1}{h} \int
(E-\mu_{L})T_{\mathrm{LR}}(E)[f_{\mathrm{L}}(E)-f_{\mathrm{R}}(E)]dE.\\
\end{split}
\end{equation}
Here, we neglect the heat current carried by the phonon, because that this part of the heat
current is usually much smaller than that induced by the electron at low temperature.
In Eq.(\ref{eq:2}),
\begin{equation}\label{eq:4}
\begin{split}
&f_{\alpha}(E,\mu_{\alpha},\mathcal{T}_{\alpha}) =\frac{1}{e^{(E-\mu_{\alpha})/k_B \mathcal{T}_{\alpha}}+1}
\end{split}
\end{equation}
is the Fermi distribution function of the left/right terminals,
where ${\alpha}=\mathrm{L}$ or $\mathrm{R}$ for the left or right terminal, and the chemical potential $\mu_{\alpha} =E_F+eV_{\alpha}$ with the Fermi energy $E_F$.

$T_{\mathrm{LR}}(E)$ in Eq.(\ref{eq:2}) is the transmission coefficient through the
3D TI nanowire. By using nonequilibrium Green's function method,
$T_{\mathrm{LR}}(E)$ can be obtained as:
$T_{\mathrm{LR}}(E)=\textmd{Tr}[{\bf \Gamma}_{\mathrm{L}}
{\bf G}^r{\bf \Gamma}_{\mathrm{R}}{\bf G}^a]$,
in which ${\bf \Gamma}_{\mathrm{L/R}}(E)=i[{\bf \Sigma}_{\mathrm{L/R}}^r(E)
-{\bf \Sigma}_{\mathrm{L/R}}^a(E)]$ and the Green's function
${\bf G}^r(E)=[{\bf G}^a]^{\dagger}=[E {\bf I}-
{\bf H}^{\mathrm{cen}}-\sum_{\alpha}{\bf \Sigma}_{\alpha}^r]^{-1}$,
with ${\bf H}^{\mathrm{cen}}$ being the Hamiltonian of center scattering region and
the self-energy ${\bf \Sigma}_{\mathrm{L/R}}^{r/a}$ stems from coupling to the left/right lead.\cite{Longw,LeeD}
For a clean TI nanowire, the center scattering region can arbitrarily be taken
and the results are exactly identical.
On the other hand, while in the presence of disorder,\cite{Chengs}
we consider that the disorder only exist in the center scattering region and
the left and right terminals are the perfect semi-infinite 3D TI nanowire still.
In the presence of disorder,
the on-site energies at the center region are added with a term $D_{nm}\sigma_{0}$ with
\begin{equation}\label{eq:3}
\begin{split}
&D_{nm} =\sum_{n',m'} \tilde{D}_{n'm'}
 \textmd{exp}\left({-\frac{r_{nm,n'm'}^{2}}{2\xi^{2}}}\right).
\end{split}
\end{equation}
Here $\tilde{D}_{n'm'}$ is uniformly distributed in the interval $[-D/2, D/2]$
with $D$ being the disorder strength,
$r_{nm,n'm'}$ is the distance between site $(n,m)$ and $(n',m')$,
and $\xi$ is the parameter describing the correlation length of the disorder.
In the numerical calculation, we consider the long range disorder with $\xi =5a$ and the
disorder density 50\%.
With each value of disorder strength $D$, the transmission coefficient $T_{\mathrm{LR}}(E)$,
the conductance, Seebeck coefficient, thermal conductance, and $ZT$
are averaged up to 40 configurations in the calculation.

In the case of very low bias and very small temperature gradient,
the Fermi distribution function in Eq.(\ref{eq:2}) can be expanded linearly
in terms of the Fermi energy $E_F$ and the temperature $\mathcal{T}$ as
\begin{equation}\label{eq:5}
\begin{split}
&f_{\mathrm{L/R}}(E,\mu_{\mathrm{L/R}},\mathcal{T}_{\mathrm{L/R}})
=f_0-eV_{\mathrm{L/R}}\frac{\partial f_0}{\partial E}+\Delta \mathcal{T}_{\mathrm{L/R}}\frac{\partial f_0}{\partial \mathcal{T}},\\
\end{split}
\end{equation}
where $\Delta \mathcal{T}_{\mathrm{L/R}} =\mathcal{T}_{\mathrm{L/R}} -\mathcal{T}$
and $f_0 =[e^{(E-E_F)/k_B \mathcal{T}}+1]^{-1}$ is the Fermi distribution function
at the zero thermal gradient and zero bias.
Then linear thermoelectric transport can be calculated while a small external bias voltage
${\delta V} =V_\mathrm{L}-V_\mathrm{R}$ or/and a small temperature gradient
${\delta \mathcal{T}} =\mathcal{T}_\mathrm{L}-\mathcal{T}_\mathrm{R}$
is applied between the left and right terminals.

By introducing the integrals $I_i (T) =\frac{1}{h}\int dE(E-E_F)^i
(-\frac{\partial f_0}{\partial E}) T_{\mathrm{LR}}(E)$ $(i=0,1,2)$,
the linear-electric conductance $G$ ($G=I_{\mathrm{L}}/\delta V$ at the zero thermal gradient),
the Seebeck coefficients $S_c$ ($S_c = \delta V /\delta \mathcal{T}$ at the zero electric current
$I_{\mathrm{L}}$ case),
and electric thermal conductance ${\kappa _e}$ (${\kappa _e} = Q_{\mathrm{L}}/\delta \mathcal{T}$
at the zero electric current)
can be expressed in very simple forms\cite{Liuj,Costi}
\begin{eqnarray}
G &= & e^{2}I_{0}(T),\label{eq:6} \\
S_c & = &-\frac{1}{e\mathcal{T}}\frac{I_{1}(T)}{I_{0}(T)},\label{eq:7}\\
\kappa_e &=&\frac{1}{\mathcal{T}}\left[I_{2}(T)-\frac{I_{1}^{2}(T)}{I_{0}(T)}\right].
\end{eqnarray}
After solving $G$, $S_c$, and $\kappa_e$, the thermoelectric figure of merit $ZT=G{{S_c}^2}\mathcal{T}/\kappa_e$
can be obtained straightforwardly.

\section{\label{sec3} thermoelectric properties at zero perpendicular magnetic field
and zero gate voltage}

\begin{figure}
\includegraphics[scale=0.34]{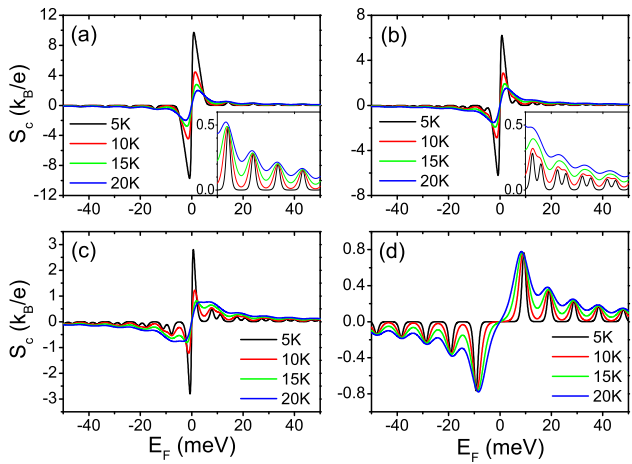}
\caption{
Panels (a)-(d) plot the Seebeck coefficients $S_c$ versus Fermi energy $E_F$ for different
temperatures with the longitudinal magnetic flux $\phi_{\parallel} = 0$, $\pi/3$,
$2\pi/3$, and $\pi$, respectively.
The insets in panels (a) and (b) are zoom-in figures of the small peaks
in the corresponding main figures.
The perpendicular magnetic field $\phi_{\perp}=0$ and gate voltage $\Delta U =0$.
}
\end{figure}

\begin{figure}
\includegraphics[scale=0.34]{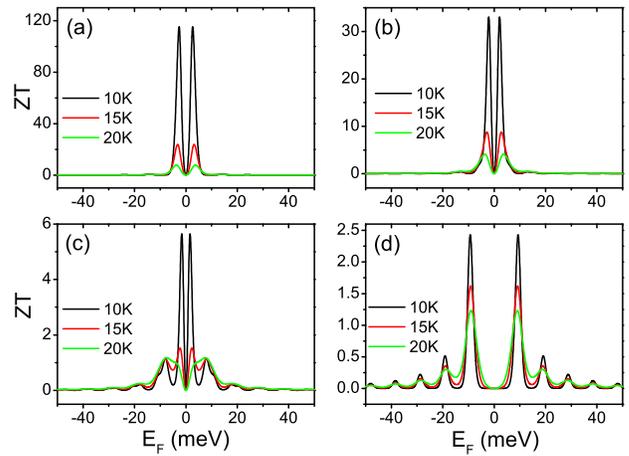}
\caption{
Panels (a)-(d) plot $ZT$ versus $E_F$ for different
temperatures with the longitudinal magnetic flux $\phi_{\parallel} = 0$, $\pi/3$,
$2\pi/3$, and $\pi$, respectively.
The perpendicular magnetic field $\phi_{\perp}=0$ and gate voltage $\Delta U =0$.
}
\end{figure}

First, we study the Seebeck coefficient $S_c$ and the thermoelectric figure of merit $ZT$
at the zero magnetic field and zero gate voltage.
Figure 2(a) and Fig.3(a) show $S_c$ and $ZT$ versus
the Fermi energy $E_F$ for different temperatures.
Due to electron-hole symmetry, $S_c$ is an odd function of the Fermi energy $E_F$ with
$S_c(-E_F) = -S_c(E_F)$. However, $ZT$ is an even function of $E_F$ with $ZT(-E_F) = ZT(E_F)$.
The properties $S_c(-E_F) = -S_c(E_F)$ and $ZT(-E_F) = ZT(E_F)$ can remain
even if in the presence of the magnetic field, gate voltage, and disorder.
$S_c$ and $ZT$ exhibit a series of peaks at low temperatures.
When $E_F$ crosses the discrete transverse channels where
the transmission coefficient $T_{\mathrm{LR}}$ jumps from one integer to another,
$S_c$ and $ZT$ show peak.
The closer the Dirac point is, the higher the peak is.
$S_c$ and $ZT$ have the highest peak near the Dirac point.
The value of $ZT$ at the highest peak exceeds over 100 at the temperature $\mathcal{T}=10 K$.
The highest peak is much higher than other peaks. For $S_c$ ($ZT$),
the highest peak is about 10 (100) times higher than the second highest peak.
This is because of the $\pi$ Berry phase around the 3D TI nanowire and the wormhole effect,
and an energy gap opens at the zero magnetic field at the Dirac point, leading that
the transmission coefficient $T_{\mathrm{LR}}=0$.
In order to balance the thermal forces acting on the charge carriers,
it needs a very large bias which results in a very large $S_c$ and $ZT$ near the Dirac point.
When the temperature rises, the height of the highest peak of $S_c$ decreases,
but the heights of the other peaks roughly remain unchanged and the valleys rise.

Next, we study the effect of the longitudinal magnetic field $B_\parallel$
on the Seebeck coefficient $S_c$ and the thermoelectric figure of merit $ZT$.
Here the longitudinal magnetic field is described by the magnetic flux
$\Phi_{\parallel}$ in the cross section of the TI nanowire,
with $\Phi_{\parallel} =L_x L_y B_{\parallel}$.
Figure 1(b) and 1(c) show the energy band structures of the TI nanowire at $\phi_{\parallel} \equiv
\Phi_{\parallel}/\phi_0 =0$ and $\pi$.
Because of a $\pi$ Berry phase for electron going around the four facets of TI nanowire,\cite{JBardarson,REgger,Zhangy}
it yields a gapped spectrum of surface state at $\phi_{\parallel} =0$.
At $\phi_{\parallel} =0$, each band is double degenerate.
With the increase of $\phi_{\parallel}$ from zero, the Aharonov-Bohm phase emerges
and the double degeneracy is removed.\cite{Dufouleur,SHong,SCho}
One sub-band moves up and other sub-band goes down, leading that the gap becomes narrower.
When the magnetic flux $\phi_{\parallel}=\pi$, the $\pi$ Aharonov-Bohm phase exactly cancels
the $\pi$ Berry phase, leading that
a pair of non-degenerate linear modes emerge with the gap closing [Fig.1(c)].
But other bands are double degenerate again.
Now it is ready to study the effect of the longitudinal magnetic field $\phi_{\parallel}$
on the thermoelectric properties.
$S_c$ and $ZT$ are the periodic functions of $\phi_{\parallel}$ with
$S_c(\phi_{\parallel}) = S_c(\phi_{\parallel}+2\pi)$
and $ZT(\phi_{\parallel}) = ZT(\phi_{\parallel}+2\pi)$.
In addition, $S_c(\phi_{\parallel}) = S_c(-\phi_{\parallel})$ and
$ZT(\phi_{\parallel}) = ZT(-\phi_{\parallel})$
because that the system is invariant by simultaneously making
the time-inversion transformation and rotation $180^\circ$ by fixing the x axis.
In Fig.2 and Fig.3, we show the Seebeck coefficient $S_c$ and the thermoelectric figure of merit
$ZT$ for the longitudinal magnetic flux
$\phi_{\parallel}=0$, $\pi/3$, $2\pi/3$, and $\pi$, respectively.
When $\phi_{\parallel}$ increases from zero, all peaks in the curves of $S_c$$\sim$$E_F$ and
$ZT$$\sim$$E_F$ split into two due to that the double degeneracy is removed.
The height of the highest peak near the Dirac point also gradually decrease.
Especially for $ZT$, the trend of decreasing is very obvious.
But the value $ZT$ is still over 30 at $\phi_{\parallel} = \pi/3$.
For $\phi_{\parallel} = \pi$, the highest peak near the Dirac point disappear completely
because of the close of the energy gap. But other peaks can remain still,
and two adjacent peaks combine into a single peak again.
In this case, $S_c$ and $ZT$ are small.
Therefore, the thermoelectric properties ($S_c$ and $ZT$) can be well adjusted
by the longitudinal magnetic field.
In fact, for $\phi_{\parallel} = \pi$, the magnetic field $B_{\parallel}$ is about 11.3 Tesla.

\begin{figure}
\includegraphics[scale=0.30]{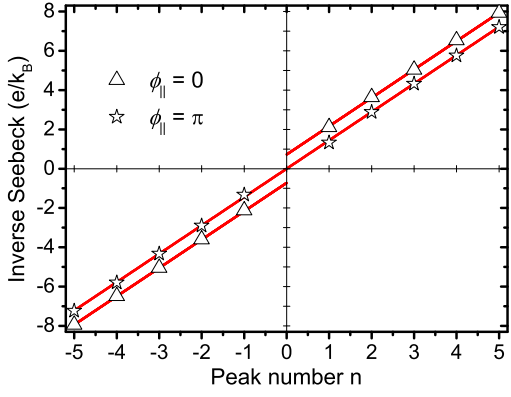}
\caption{ The inverse of peak height of Seebeck coefficients $S_c$ vs
the peak number $n$.
The upper triangle symbol and hollow pentagram symbol denote the magnetic flux $\phi_{\parallel}= 0$
and $\pi$. These data points are obtained from the curve of $\mathcal{T} =5K$ in Fig.2(a) and 2(d).
The two red lines are $\frac{e}{k_B} (n + \texttt{sign}(n)/2)/\texttt{ln}2$ and
$\frac{e}{k_B} n/\texttt{ln}2$.
}
\end{figure}

Figure 4 shows the inverse of the peak height of Seebeck coefficient $S_c$
versus the peak number $n$ with the longitudinal magnetic flux $\phi_{\parallel}=0$ and $\pi$.
Here the peak number $n$ denotes the $n$-th peak near the Dirac point,
but the highest peak does not count at $\phi_{\parallel}=0$.
In fact, except for the highest peak,
the heights of the other peaks are almost independent of temperature [see Fig.2(a) and 2(d)].
We can see that at $\phi_{\parallel} =0$
the inverse of the peak height is proportional to $\frac{e}{k_B} (n + \texttt{sign}(n)/2)/\texttt{ln}2$
with $\texttt{sign}(n) =1$ for $n>0$ and $-1$ for $n<0$ (see the upper triangle symbol in Fig.4).
This is similar as that in the conventional metal.\cite{Xingy}
On the other hand, for $\phi_{\parallel} =\pi$,
the inverse of the peak height is proportional to $\frac{e}{k_B} n/\texttt{ln}2$
(see the hollow pentagram symbol in Fig.4), which is similar as that in graphene.\cite{Xingy}
This is because that the extra $\pi$ Berry phase and the wormhole effect
lead to a half-integer shift in the curve of the inverse of the peak height of $S_c$ versus the peak number
$n$.
In fact, these conclusions can also analytically be obtained from the energy band
structure and the transmission coefficient $T_{\mathrm{LR}}(E)$.
Taking $\phi_{\parallel} =0$ with the positive $n$ as an example,
when the energy $E$ is in the vicinity of $E_n$, the transmission coefficient $T_{\mathrm{LR}}(E)$
can be written as $T_{\mathrm{LR}}(E) = 2n$ at $E< E_n$ and it jumps to $2n+2$ at $E>E_n$
with $E_n$ being the bottom of the $n$-th sub-band.
Then substituting this transmission coefficient $T_{\mathrm{LR}}(E)$ into Eq.(\ref{eq:7}),
we can obtain
\begin{equation}\label{eqshape}
S_c (E_F) = \frac{k_B}{e} \frac{-x e^x + (1+e^x)\texttt{ln} (1+e^x)}
          {1+ n(1+e^x)},
\end{equation}
with $x \equiv (E_n -E_F)/k_B \mathrm{T}$.
This equation gives the shape of the $n$-th peak for $\phi_{\parallel} =0$ with the positive $n$.
So the height of the $n$-th peak of $S_c$ is about $\frac{k_B}{e} \texttt{ln}2 /(n + 1/2)$.
From $S_c (-E_F) =-S_c (E_F)$, the peak heights for the negative $n$ can be obtained as
$\frac{k_B}{e} \texttt{ln}2 /(n - 1/2)$ straightforwardly.
Similarly, the shape of the $n$-th peak of $S_c$ for $\phi_{\parallel} =\pi$
can analytically be derived
\begin{equation}\label{eqshape}
S_c (E_F) = \frac{k_B}{e} \frac{-x e^x + (1+e^x)\texttt{ln} (1+e^x)}
          {1+ (n-1/2)(1+e^x)},
\end{equation}
and the peak height is $\frac{k_B}{e} \texttt{ln}2 /n$.
In Fig.4, the curves $\frac{e}{k_B} (n + \texttt{sign}(n)/2)/\texttt{ln}2$ and
$\frac{e}{k_B} n/\texttt{ln}2$ (the analytic results for the inverse of the peak height of $S_c$)
are also shown. They are well consistent with the numerical points.

\begin{figure}
\includegraphics[scale=0.32]{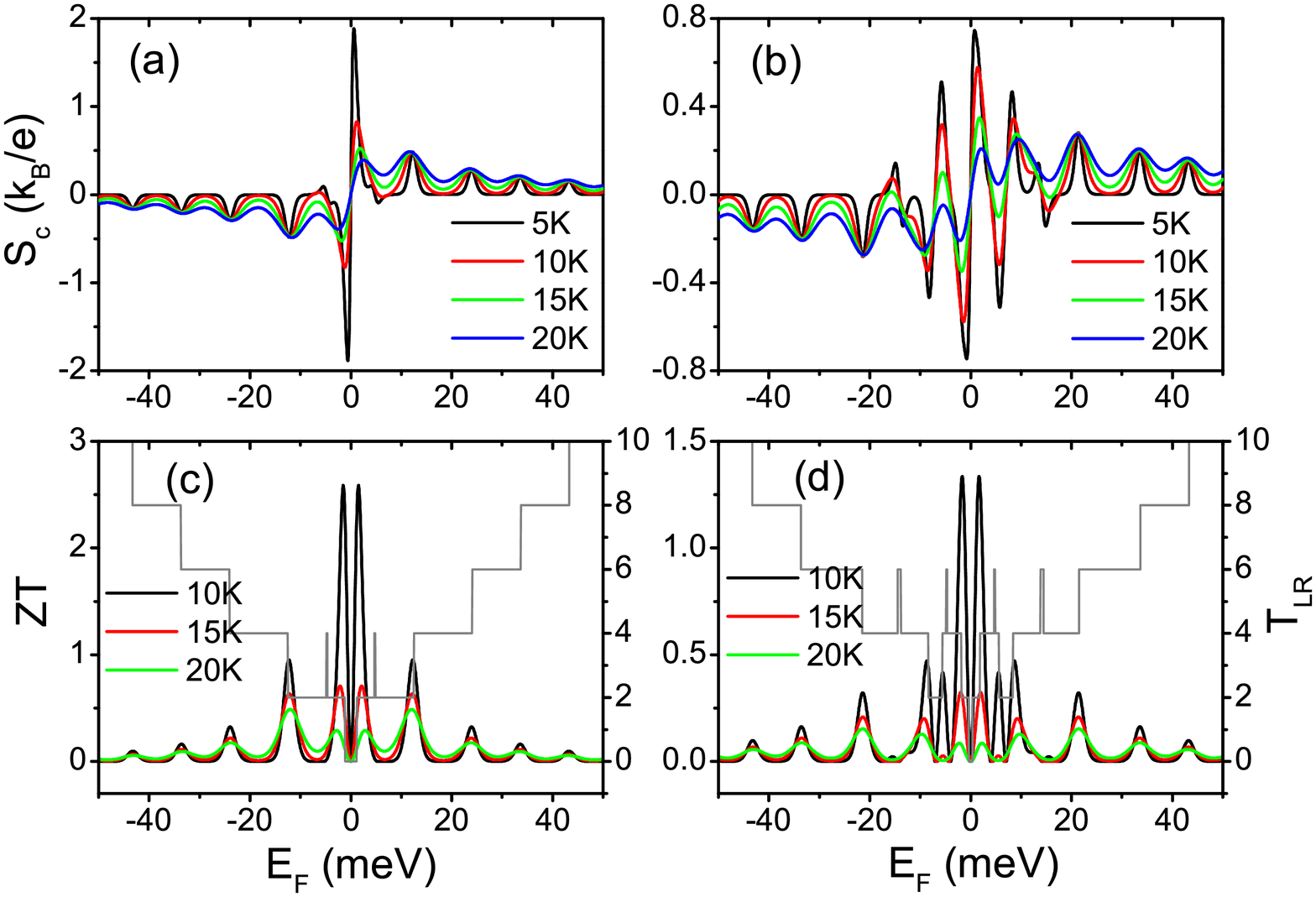}
\caption{ The Seebeck coefficient $S_c$ and $ZT$ vs Fermi energy $E_F$
for different temperatures. $\phi_{\parallel} = 0$, $\phi_{\perp} = 0$,
and the gate voltage $\Delta U = 30$meV in panels (a) and (c) and $\Delta U = 50$meV in panels (b) and (d).
The gray curves in (c) and (d) are the transmission coefficient $T_{\mathrm{LR}}$.
}
\end{figure}

\begin{figure}
\includegraphics[scale=0.32]{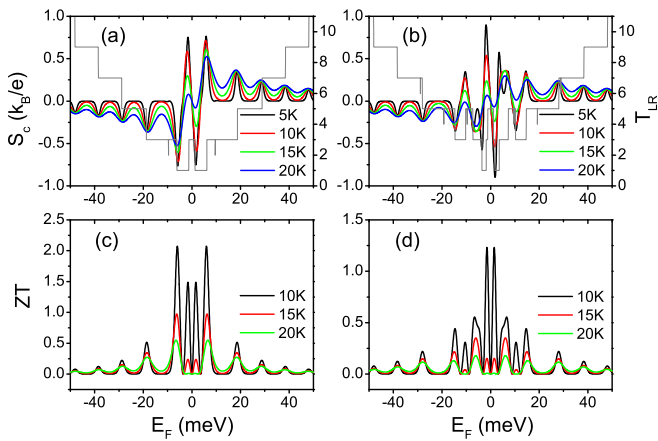}
\caption{ $S_c$ and $ZT$ vs Fermi energy $E_F$
for different temperatures. $\phi_{\parallel} = \pi$, $\phi_{\perp} = 0$,
and the gate voltage $\Delta U = 30$meV in panels (a) and (c) and $\Delta U = 50$meV in panels (b) and (d).
The gray curves in (a) and (b) are the transmission coefficient $T_{\mathrm{LR}}$.
}
\end{figure}

\section{\label{sec4} effect of the perpendicular magnetic field and gate voltage on
thermoelectric properties}

In this section, we study the effect of the perpendicular magnetic field $B_{\perp}$
and gate voltage $\Delta U$ on the Seebeck coefficient $S_c$ and
thermoelectrical figure of merit $ZT$ of the 3D TI nanowire.
First, the effect of $\Delta U$ is studied.
Figure 5 shows $S_c$ and $ZT$ with $\phi_{\parallel} = 0$, and the gate voltage $\Delta U$ being 30meV and 50meV.
The parameters in Fig.6 are similar to Fig.5, but $\phi_{\parallel} = \pi$.
In order to explain the behavior of $S_c$ and $ZT$ clearly,
the transmission coefficient $T_{\mathrm{LR}}$ is also given in Fig.5 and Fig.6,
and here $T_{\mathrm{LR}}$ is quantized and exhibits a series of plateaus.
In Fig.5, in which the longitudinal magnetic flux $\phi_{\parallel}=0$,
we see that $S_c$ and $ZT$ show peaks when $T_{\mathrm{LR}}$ jumps from one step to another.
In particular, as the gate voltage increases, the large $ZT$ at $\Delta U=0$ [see Fig.3(a)]
reduces swiftly. When $\Delta U =20$meV, the value of $ZT$ is about $15.6$.
So the gate voltage can regulate the thermoelectric properties.
In addition, the oscillation peak near the Dirac point becomes
dense in the presence of $\Delta U$ [see Fig.5(b) and (d)].
Because that the gate voltage $\Delta U$ causes the difference between the potential energies
of the upper and lower surfaces, it affects the states of the
side surfaces, and makes the energy band deform, which leads to a strong reduction of the energy gap.
Figure 6 shows the curve of $S_c$$\sim$$E_F$ and $ZT$$\sim$$E_F$
at the longitudinal magnetic flux $\phi_{\parallel} =\pi$.
Similarly, $S_c$ and $ZT$ display peaks when the transmission coefficient $T_{\mathrm{LR}}$
steps jump.
The dense peaks are also displayed near Dirac point when the gate voltage $\Delta U$ increases.
By comparing between Fig.5(b,d) and Fig.6(b,d), for $\Delta U$ = 50meV,
we find that the curves of $S_c$$\sim$$E_F$ and $ZT$$\sim$$E_F$ are very similar,
although $\phi_{\parallel}=0$ in Fig.5 and $\phi_{\parallel}=\pi$ in Fig.6.
This means that both the $\pi$ Berry phase and the Aharonov-Bohm phase of $\phi_{\parallel}$
have little effect on the Seebeck coefficient $S_c$ and $ZT$ while at the large gate voltage.

\begin{figure}
\includegraphics[scale=0.32]{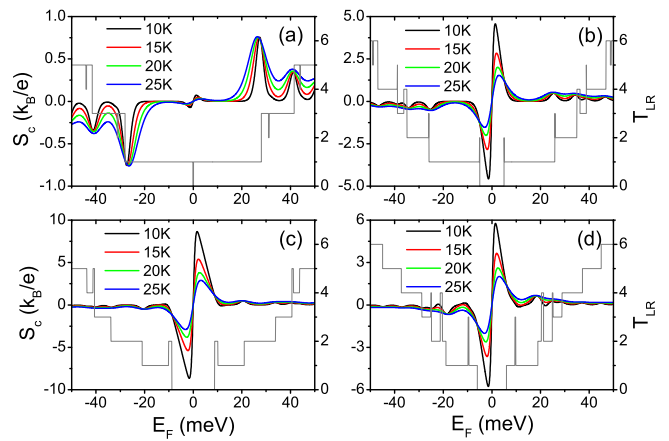}
\caption{ The Seebeck coefficient $S_c$ vs $E_F$ for different temperatures
with the perpendicular magnetic field $\phi_{\perp} = 0.005$ and $\phi_{\parallel}=0$.
The gate voltage $\Delta U = 0$meV (a), $10$meV (b), $20$meV (c), and $50$meV (d).
The gray curves in (a-d) are the transmission coefficient $T_{\mathrm{LR}}$.
}
\end{figure}

\begin{figure}
\includegraphics[scale=0.34]{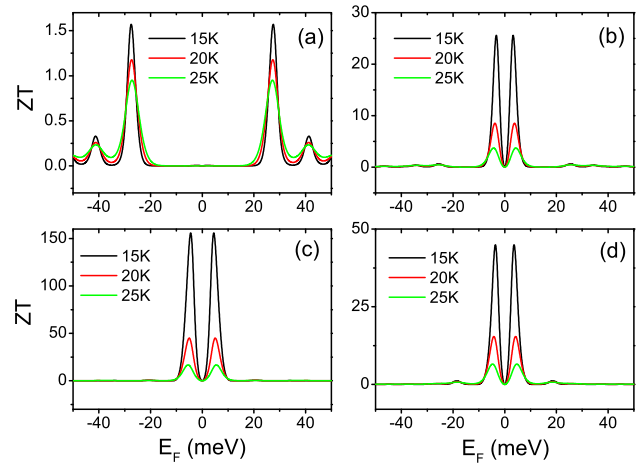}
\caption{ $ZT$ vs $E_F$ for different temperatures
with $\phi_{\perp} = 0.005$ and $\phi_{\parallel}=0$.
The gate voltage $\Delta U = 0$meV (a), $10$meV (b), $20$meV (c), and $50$meV (d).
}
\end{figure}

Now, we study the effect of the perpendicular magnetic field $B_{\perp}$ on
the Seebeck coefficient $S_c$ and the thermoelectric figure of merit $ZT$.
For the small $B_{\perp}$ (e.g. $B_{\perp}< 0.1$ Tesla), $S_c$ and $ZT$ are are almost unaffected,
and $ZT$ has still large value at $\phi_{\parallel}=0$.
On the other hand, for a large perpendicular magnetic field $B_{\perp}$,
Landau levels form and edge states appear on the side surfaces.
In this case, $S_c$ and $ZT$ are almost independent of the
longitudinal magnetic field $\phi_{\parallel}$ and $ZT$ strongly reduces.
Figure 7(a) and Fig.8(a) show $S_c$ and $ZT$ versus the Fermi energy $E_F$
with the perpendicular magnetic flux in a lattice $\phi_{\perp} = 0.005$
(the real magnetic field $B_{\perp}$ is around 18.3 Tesla).
The $S_c$ displays peaks when $E_F$ passes the Landau levels
and show valleys between adjacent Landau levels.
Because that the Landau levels are highly degenerate, the number of energy levels decreases,
and as a result the peak spacing becomes larger and the peak becomes sparse.
When $E_F$ is on the zeroth Landau level, $S_c$ is zero.
This is because the zeroth Landau level with the doubly degeneracy is
shared equally by electrons and holes,
and the electrons and holes give the opposite contributions to $S_c$.
Moreover, the peak height of $S_c$ is proportional to
$\frac{k_B}{e}\texttt{ln}2/n$ with the peak number $n$.
With the increase of temperature $\mathcal{T}$, the peak height of $S_c$ remains approximately
unchanged, but the valley rises, which shows that $S_c$ peaks are robust against the temperature.
For the thermoelectric figure of merit $ZT$, it is small for all the Fermi energy $E_F$,
because of the appearance of the edge states and the absence of the energy gap at the large $\phi_{\perp}$.
In addition, there are two largest peaks in $ZT$
at $E_F \approx \pm 27.3$meV (27.3meV is the first Landau level).
The positions of the $ZT$ peaks are corresponding to the $S_c$ peaks.
With the increase of $\phi_{\perp}$, the peak spacing of $ZT$ becomes larger and
the peak becomes sparse similar to the peaks of $S_c$.

Let us study the case of the coexistence of both the perpendicular magnetic field $\phi_{\perp}$
and gate voltage $\Delta U$.
Figure 7(b-d) and Fig.8(b-d) show $S_c$ and $ZT$ at the large $\phi_{\perp}$
($\phi_{\perp}=0.005$) and zero $\phi_{\parallel}$ for the different $\Delta U$.
For the large $\phi_{\perp}$, the Landau levels form, and both $S_c$ and $ZT$
are almost independent of the longitudinal magnetic flux $\phi_{\parallel}$.
When the gate voltage $\Delta U$ is applied, the Landau levels of the upper and lower surfaces
split, and then it produces a gap spectrum of surface states in the TI nanowire.
So the highest peaks of $S_c$ and $ZT$ near the Dirac point appear,
and the value of $ZT$ strongly increases. While $\Delta U \geq 10$meV, $ZT$ can
exceed over $25$.
In addition, we can see from Fig.7 and Fig.8 that when the gate voltage increases,
the bandgap of surface states increases first and then decreases due to the side surfaces.
So the highest peak of $S_c$ and $ZT$ also tends to increase first and then decrease.
But $ZT$ can remain the large value in a very large range of $\Delta U$.
In short, by adjusting the longitudinal, perpendicular magnetic fields and gate
voltage, it is easy to change the value of $S_c$ and $ZT$ greatly,
i.e. to change greatly the thermoelectric properties of the 3D TI nanowire.

\section{\label{sec5} effect of the disorder on
thermoelectric properties }

\begin{figure}
\includegraphics[scale=0.34]{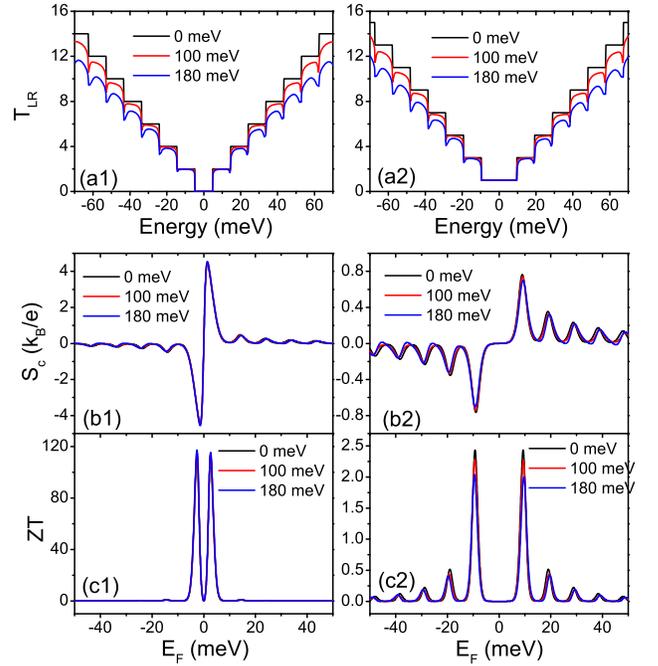}
\caption{ The transmission coefficient $T_{\mathrm{LR}}$ vs the energy $E$ (a), the Seebeck coefficient $S_c$ vs $E_F$ (b)
and $ZT$ vs $E_F$ (c) for the different disorder strengths $D$.
The temperature $\mathcal{T} =10K$, the gate voltage $\Delta U =0$ meV, $\phi_{\perp}=0$,
and $\phi_{\parallel}=0$ for panels (a1)-(c1) and $\pi$ for panels (a2)-(c2). }
\end{figure}

\begin{figure}
\includegraphics[scale=0.34]{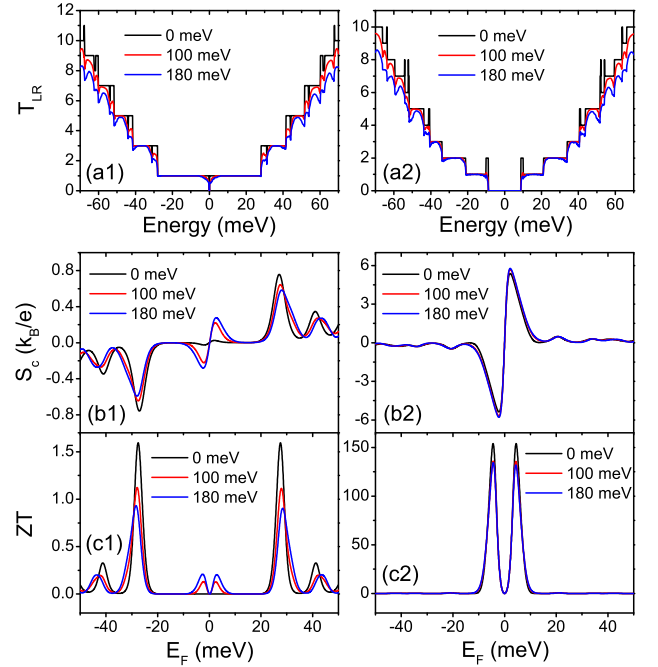}
\caption{ $T_{\mathrm{LR}}$ vs energy $E$ (a), $S_c$ vs $E_F$ (b) and
$ZT$ vs $E_F$ (c) for the different disorder strengths $D$.
The perpendicular magnetic field $\phi_{\perp} = 0.005$, the longitudinal magnetic field
$\phi_{\parallel}=0$, the temperature $\mathcal{T} =15K$,
and the gate voltage $\Delta U = 0$meV for the panels (a1)-(c1) and $20$ meV for the panels (a2)-(c2).
}
\end{figure}

Up to now, we have shown that the Seebeck coefficient $S_c$ and the thermoelectric figure
of merit $ZT$ have the large value at zero magnetic fields with the zero gate voltage, or at the large
perpendicular magnetic field with the nonzero gate voltage.
Next, let us study the effect of the disorder on $S_c$ and $ZT$.
Figure 9 shows the transmission coefficient $T_{\mathrm{LR}}$,
$S_c$ and $ZT$ for the different disorder strength $D$ at the zero perpendicular magnetic field.
When the disorder strength $D =0$, $T_{\mathrm{LR}}$ displays quantum plateaus.
While in the presence of the disorder ($D\not=0$),
the lower plateaus (e.g. the plateaus with $T_{\mathrm{LR}}=0$ and $1$) are robust against disorder.
On the other hand, the higher plateaus of $T_{\mathrm{LR}}$ are obviously destroyed,
because that the scattering occurs by the disorders.
But the results show that $S_c$ and $ZT$ are very robust against disorder
in a wide range of Fermi energy $E_F$.
Not only the highest peak near Dirac point can survive, but also the low peak at large $E_F$
can remain at the strong disorder.
Even the disorder strength $D=180$meV, the peak value of $ZT$ can still exceed over 100
[see Fig.9(c1)] and these lower peaks remain almost the same [see Fig.9(b1), (b2) and (c2)].
This is obviously different from the intuition,
because thermoelectric behaviors ($S_c$ and $ZT$) are more sensitive to the density of state
than the conductance (or transmission coefficient).
In fact, although the plateaus of $T_{\mathrm{LR}}$ are obviously destroyed by the disorder,
the sudden jumps from one plateau to another still exist
and the position of the jump point remains unchanged.
Because the peaks of $S_c$ and $ZT$ are mainly determined by the jumps
of $T_{\mathrm{LR}}$, they are robust against the disorder.
The characteristics of $ZT$ having the large value and being robust against the disorder
are beneficial for the application of the thermoelectricity.

Figure 10 shows the transmission coefficient $T_{\mathrm{LR}}$, $S_c$ and $ZT$
for the different disorder strength at the high
perpendicular magnetic field $\phi_{\perp} = 0.005$ with the gate voltage $\Delta U=0$ and
$20$meV.
While $\Delta U=0$, $S_c$ and $ZT$ are small for the clean TI nanowire.
The disorder reduce the heights of the peaks of $S_c$ and $ZT$,
and slightly shifts the peak positions also.
For example, while the disorder strength $D=180$meV,
the peak heights of $ZT$ decrease to about half of that at $D=0$.
On the other hand, for the case with the non-zero gate voltage (e.g. $\Delta U =20$meV),
$S_c$ and $ZT$ have the large peak values while the disorder strength $D=0$.
With the increasing of $D$, the peak heights and positions of $S_c$ and $ZT$
can remain unchanged almost.
When $D=180$meV, the largest value of $ZT$ can still exceed over 100.
$ZT$ not only has a large value, but also is robust against the disorder,
which is very promising for the application.

\section{\label{sec6}Conclusions}

In summary, we study the magnetothermoelectric transport properties of the surface states of
3D TI nanowires under
the longitudinal and perpendicular magnetic fields.
The Seebeck coefficient $S_c$ and the thermoelectric figure of merit $ZT$ show peaks
where there are step changes of transmission coefficient.
Due to the electron-hole symmetry, the Seebeck coefficient is odd function of the Fermi energy $E_F$,
and $ZT$ is even function.
The highest peak appears when $E_F$ is near the Dirac point, and
the peak heights gradually decrease with $E_F$ far from the Dirac point.
At the zero magnetic field and zero gate voltage, the Seebeck coefficient and $ZT$
have the large peak value due to the $\pi$ Berry phase around the topological insulator
nanowire and the wormhole effect.
The Seebeck coefficient $S_c$ and $ZT$ are obviously dependent on
the gate voltage, the longitudinal, and perpendicular magnetic fields.
This means that the thermoelectric properties of the 3D TI nanowire
can be easily adjusted by tuning the gate voltage or magnetic fields.
At zero magnetic fields and zero gate voltage, or at the large
perpendicular magnetic field and nonzero gate voltage,
$ZT$ has the large value.
In addition, the effect of the disorder on the thermoelectric properties is also studied.
It is a surprise that the Seebeck coefficient and $ZT$ are more undisturbed
than the conductance (transmission coefficient). The plateaus of transmission coefficient
can be broken by the disorder, but the peaks at the Seebeck coefficient and $ZT$ are robust
against the disorder, because the jumps of transmission coefficient can remain in the presence
of the disorder.
The characteristics, that $ZT$ has the large value and is robust against the disorder,
are very beneficial for the application of the thermoelectricity.

\section*{Acknowledgement}
This work was financially supported by National Key R and D Program of China (2017YFA0303301),
NBRP of China (2015CB921102), NSF-China (Grants No. 11574007), and
the Key Research Program of the Chinese Academy of Sciences (Grant No. XDPB08-4).

\end{document}